\documentstyle[aps,multicol,epsfig]{revtex}

\begin{document}
\title{Derivative moments in turbulent shear flows}
\author{J. Schumacher$^{1}$, K.R. Sreenivasan$^2$ and P.K. Yeung$^3$}
\address{$^1$Fachbereich Physik, Philipps-Universit\"at, D-35032 Marburg,
Germany\\
$^2$Institute for Physical Science and Technology, University of
Maryland, College Park, MD 20742\\
$^3$School of Aerospace Engineering, Georgia Institute of
Technology, Atlanta, GA 30332}
\date{\today}
\maketitle

\begin{abstract}
We propose a generalized perspective on the behavior of high-order
derivative moments in turbulent shear flows by taking account of
the roles of small-scale intermittency and mean shear, in addition
to the Reynolds number. Two asymptotic regimes are discussed with
respect to shear effects. By these means, some existing
disagreements on the Reynolds number dependence of derivative
moments can be explained. That odd-order moments of transverse
velocity derivatives tend not vanish as expected from elementary
scaling considerations does not necessarily imply that small-scale
anisotropy persists at all Reynolds numbers.\\

\end{abstract}

\begin{multicols}{2}
\section{Introduction}
The postulate of local isotropy \cite{Kolmogorov41} implies an
invariance with respect to spatial rotations of the {\it
statistical} properties of small scales of turbulence. Even though
the large scales are anisotropic in all practical flows, it is
thought that the small scales at high Reynolds numbers are
shielded from anisotropy because of their separation by a wide
range of intermediate scales. At any finite Reynolds number, some
residual effects of small-scale anisotropy may linger, but all
proper measures of anisotropy are expected to decrease rapidly
with Reynolds number. An understanding of the rate at which small
scales tend towards isotropy is a basic building block in
turbulence theory.

A particularly appealing manner of generating large-scale
anisotropy is by a homogeneous shear characterized by a constant
shear rate $S \equiv dU/dy$, where $U(y)$ is the mean velocity in
the streamwise direction $x$, and $y$ is the direction of the
shear. During the last few years, nearly homogeneous shear flows,
both experimental \cite{Garg98,Ferchichi00,Shen00} and numerical,
\cite{Pumir96,Schu00,Schu01} have examined the rate at which local
isotropy is recovered with respect to the Taylor microscale
Reynolds number, $R_{\lambda}$. The notation is standard:
$R_\lambda \equiv u^{\prime} \lambda/\nu$, $u$ is the velocity
fluctuation in the longitudinal direction $x$, $u^{\prime} \equiv
\sqrt{\langle u^2 \rangle}$, $\lambda^2 = \langle u^2
\rangle/\langle (\partial u/\partial x)^2 \rangle$, $\nu$ is the
kinematic viscosity and $\langle\cdot\rangle$ denotes a suitably
defined average. The discussion has often been focused on the
behavior of normalized odd moments of transverse velocity
derivatives defined as
\begin{eqnarray}
M_{2n+1}(\partial u/\partial y)= \frac{\langle (\partial
u/\partial y)^{2n+1} \rangle}
     {\langle (\partial u/\partial y)^2 \rangle^{(2n+1)/2}}\,,
\end{eqnarray}
where $n$ is a positive integer. The velocity derivatives are
small-scale quantities, and symmetry considerations of local
isotropy demand that the odd moments of transverse velocity
derivatives be zero. In practice, they should decrease with
$R_{\lambda}$ relatively rapidly. Though the postulate of local
isotropy does not by itself stipulate this rate, simple estimates
can be made by retaining the spirit of local isotropy and making
further assumptions. Let us assume that the non-dimensional
moments $M_{2n+1}$ depend on some power $p$ of the shear, which is
the ultimate source of anisotropy in homogeneous shear flows, and
non-dimensionalize the shear dependence by a time scale formed by
the energy dissipation rate $\epsilon$ and fluid viscosity $\nu$;
this non-dimensionalization accords with Kolmogorov's first
hypothesis that small-scale properties depend solely on $\epsilon$
and $\nu$. We may then write
\begin{equation}
M_{2n+1}(\partial u/\partial y)= S^p f(\epsilon, \nu) \sim
R_{\lambda}^{-p}, \label{lumley}
\end{equation}
where it is further assumed that $S = {\cal O}(u^\prime/L)$ and
$L$ is the large scale of turbulence. Lumley \cite{Lumley67}
considered a linear dependence on the shear (i.e., $p = 1$). This
yields an inverse power of $R_{\lambda}$ for the decay of all odd
moments. The choice $p = 1$ accounts for the dependence of the
sign of the odd moments on the sign of $S$ in a simple manner.

The existing experimental and numerical data on the skewness of
the transverse velocity derivative are collected in Fig.~1. Data
from any given source seem to fit a power-law of the form
\begin{eqnarray}
\label{sceq1} M_3(\partial u/\partial y)\sim R_{\lambda}^{-m}\,
\end{eqnarray}
with $m$ differing from one set of data to another. Though the
power-law is not a particularly good fit for the totality of the
data, the average roll-off seems to be less steep than the inverse
power discussed above. Atmospheric data at much higher Reynolds
numbers \cite{Dhruva98,Kurien00} are consistent with this slower
rate of decay.

The situation with respect to the hyperksewness, $M_5$, is as
follows. Two independent laboratory experiments in nearly
homogeneous shear flows \cite{Ferchichi00,Shen00} draw different
conclusions on the $R_{\lambda}$ behavior of $M_5$. On the one
hand, Shen and Warhaft \cite{Shen00} find no dependence on
$R_{\lambda}$ in the range between $10^2$ and $10^3$. On the other
hand, Ferchichi and Tavoularis \cite{Ferchichi00} regard their
data to be essentially consistent with expectations of local
isotropy. (For one perspective on this difference, see Warhaft and
Shen \cite{Warhaft01}). A collection of all known data is given in
Fig.~2. The overall impression from the figure is that, while
there is probably a weak decreasing trend for $R_{\lambda} > 300$,
the hyperskewness does not diminish perceptibly even when
$R_\lambda$ is as high as 1000. Shen and Warhaft \cite{Shen00}
measured the seventh normalized moment of the velocity derivative
and found that it increased with $R_\lambda$ instead of
decreasing.

Such findings have been interpreted (e.g., Refs.~4 and 5) to mean
that, in the presence of large-scale shear, small-scale anisotropy
persists to very high Reynolds numbers. The intent of this paper
is to clarify, at least partially, the simultaneous role played by
shear, intermittency and the Reynolds number---all of which have
an impact on trends displayed by odd derivative moments.

We now consider in Sec.~II the issue of intermittency versus
anisotropy. In Sec.~III, we highlight the effects of shear by
discussing the limiting cases of large shear and local isotropy,
and argue that $R_{\lambda}$ determines the state of the flow only
partially. In Sec.~IV the derivative skewness data are plotted in
the plane spanned by the non-dimensional shear parameter $S^*$ and
$R_{\lambda}$, where we use the definition
\begin{equation}
S^*=S u^{\prime 2}/\epsilon.
\end{equation}
This broader perspective may help resolve some seemingly
contradictory claims on the recovery of isotropy of small scales.

\section{Intermittency versus anisotropy}
A large number of measurements have shown convincingly that
high-order $even$ moments of small-scale features of turbulence
increase with $R_{\lambda}$. Consider the longitudinal velocity
derivative $\partial u/\partial x$. The product of its second
moment and fluid viscosity is essentially the energy dissipation,
which is known to remain independent of $R_\lambda$ when
$R_\lambda$ exceeds some moderately high value (see Refs.~14,15).
On the other hand, all high-order moments of longitudinal velocity
gradients grow with $R_\lambda$ (see, for example, the compilation
in Ref.~16 of the data on the flatness factor of $\partial
u/\partial x$). Sixth and higher order moments increase at
increasingly faster rates with $R_\lambda$. These growths are
attributed to the intermittency of small-scale turbulence. At the
present level of our understanding, intermittency is independent
of anisotropy effects. Therefore, just as the growth of high-order
even moments with increasing $R_\lambda$ is unrelated to
anisotropy, it is legitimate to ask if, at least in part, the
slower-than-expected decay---or even modest growth---of odd
moments, may be related to intermittency.

To separate intermittency effects from those of anisotropy, at
least in some approximate way, it is useful to plot the ratio
$-M_{2n+1}(\partial u/\partial y)/M_{2n+1}(\partial u/\partial
x)$. It is plausible to assume that the $R_\lambda$-growth due to
intermittency effects is the same for the moments of $\partial
u/\partial x$ and for the moments of $\partial u/\partial y$, the
intermittency effects of odd moments of $\partial u/\partial y$
are cancelled in these ratios by those of $\partial u/\partial x$.
Though this not a rigorous statement, it is useful to see the
outcome. Figure 3 shows the results. It is clear, despite large
scatter, that all the moments show a tendency to diminish with
Reynolds number. The odd-moments in Fig.~3 are normalized by
powers of their respective variances. It would have been desirable
to plot ratios of unnormalized, or ``bare", moments of the two
derivatives, but Ref.~4 does not include those data. In any case,
this should not make much difference because $\langle(
\partial u/\partial x)^2\rangle/\langle(\partial u/\partial y)^2
\rangle$ is essentially a constant at high Reynolds numbers.

This same issue can be rephrased and reexamined in a somewhat
different light. When we consider the moments such as skewness and
hyperskewness, we usually normalize them by the appropriate power
of the variance of the variable. This is perfectly reasonable for
Gaussian or near-Gaussian variables, but not so for intermittent
quantities with highly stretched tails. Perhaps a more reasonable
alternative is to consider how an odd moment of a certain order
varies with respect to the even moment just below or just above,
or the geometric mean of those just below and just above. We
illustrate the results of this consideration for the third, fifth
and seventh order moments of $\partial u/\partial y~(=x)$ in
Fig.~4. The lack of data on the eight moment of $\partial
u/\partial y$ makes the analysis of the seventh moment incomplete.
Nevertheless, it is clear that all these alternative ways of
normalization show substantial decay. It is hard to be precise
about the rates of decay, partly because of the large scatter and
partly because the incomplete manner in which the seventh moments
have been analyzed, but it is conceivable that increasingly
high-order moments, within a given normalizing scheme, decay more
slowly. At the least, a careful discussion of the restoration of
anisotropy requires the proper inclusion of intermittency effects.
This is our first point.

\section{Shear effects in two limiting cases}
It is reasonable to suppose that, to a first approximation, the
mean shear $S$ and the viscosity $\nu$ determine the gross state
of the flow. Expressing time units in terms of $S^{-1}$, length
units in terms of the integral length scale $L$, and the mean
profile as ${\bf U}=S y {\bf e}_x$, we get
\begin{eqnarray}
\label{nseq}
\frac{\partial{\bf v}}{\partial t}+({\bf v}\cdot{\bf \nabla}){\bf v}+
Sy\frac{\partial{\bf v}}{\partial x}+ S v {\bf e}_x
&=&-{\bf \nabla} p+\frac{1}{Re_s}{\bf \nabla}^2{\bf v}+{\bf f}\;,\\
\label{ceq}
{\bf \nabla}\cdot{\bf v}&=&0\;,
\end{eqnarray}
where ${\bf v}=(u,v,w)$ and ${\bf e}_x$ is unit vector in the
streamwise direction $x$. The volume forcing is denoted by ${\bf
f}$. The two parameters may be expected to set the steady state
fluctuation level and energy dissipation rate. It also follows
that the derived parameters $R_{\lambda}$ and $S^*$ adjust
themselves dynamically, in ways that are understood only
partially, to the imposed values of $\nu$ and $S$. We then expect
\begin{eqnarray}
R_{\lambda}&=&g_1(\nu, S)=g_1(Re_s)\,,\nonumber\\
S^{\ast}&=&g_2(\nu, S)=g_2(Re_s)\,,
\end{eqnarray}
where $g_1$ and $g_2$ are unknown functions of their arguments.

All the homogeneous shear experiments to be discussed below have
been done in air. This fixes the viscosity to be approximately
constant, so we can test the dependence of $S^*$ and $R_{\lambda}$
on the shear rate $S$. The relevant plots (Figs.~5~(a) and (b))
show no obvious trend but only large scatter. This scatter may be
related in part to the fact that $P/\epsilon\ne 1$ in some of the
experiments, leading to nonstationarity. Here $P$ is the
production of turbulent kinetic energy, defined as $P=-\langle u
v\rangle S$ for homogeneous shear. In part, it demonstrates that
the flow might depend additionally on initial conditions, $\langle
{\bf v}^2 \rangle_{t=0}$, or the type of driving of small-scale
fluctuations characterized globally by an input energy. From
Eq.~(\ref{nseq}), the latter is given by
\begin{eqnarray}
\epsilon_{in}=\langle {\bf v\cdot f}\rangle\,.
\end{eqnarray}
We then get more complex relations such as
\begin{eqnarray}
R_{\lambda}&=&\tilde{g}_1(\nu, S, \epsilon_{in})=
              \tilde{g}_1(Re_s, \epsilon_{in})\,,\nonumber\\
S^{\ast}&=&\tilde{g}_2(\nu, S, \epsilon_{in})=
           \tilde{g}_2(Re_s, \epsilon_{in})\,.
\end{eqnarray}

If we add to Figs.~5 the atmospheric data from Refs.~9 and 10, the
situation becomes even more complex. However, it is likely that,
in such inhomogeneous flows, one has to take into account
secondary factors such as convective effects (though conditions in
which they are modest can always be chosen carefully). In
laboratory experiments, secondary effects might arise from the use
of specific passive or active grids for the generation of
turbulence. Further differences can arise when measuring at a
fixed point instead of following the downstream evolution. This
discussion merely underlines the inadequacy of $R_{\lambda}$ as
the sole indicator of the state of the flow. At the least, we have
to possess some knowledge of the other parameters influencing the
state of the flow before drawing firm conclusions on the recovery
of isotropy.

To keep matters simple, we will focus below on homogeneously
sheared flows. Because the initial conditions are not known in all
quantifiable details, we shall tentatively stipulate a simple
generalization of Eq.~(1) in the form
\begin{eqnarray}
S_3=\tilde{f}(R_{\lambda}, S^{\ast})\,,
\end{eqnarray}
and regard other effects as superimposed ``noise''
\cite{Gualtieri}. If so, we should investigate the behavior of the
derivative moments by keeping one of the two parameters fixed
while varying the other, for example by fixing $S^*$ and varying
$R_{\lambda}$. This is the topic of the next section. However, it
is useful to preface this consideration by examining two limiting
behaviors in which some inequalities between $R_{\lambda}$ and
$S^*$ can be established.

\subsection{Large shear case}
Consider the case of large shear rate for which the coupling of
the mean shear to the small-scale flow dominates. In the rapid
distortion limit, namely $S\to\infty$, Eq.~(\ref{nseq}) becomes
linear because the viscous term as well as the $({\bf v}\cdot{\bf
\nabla}){\bf v}$ term can be dropped, so that any shear rate
dependence can be eliminated by the rescaling of the variables,
e.g. $t\to St$.

Our dimensional estimates are related to large but finite shear
rates. For this case, the term representing the coupling of the
turbulent velocity component to the mean shear is important and
large in comparison to the nonlinear advective transport. Our
situation corresponds to the case in which
\begin{eqnarray}
\left| v_i \frac{\partial U_j}{\partial x_i}\right| \gg
\left| v_i \frac{\partial v_j}{\partial x_i}\right| \,.
\end{eqnarray}
For the homogeneous shear flow, we get for the left-hand-side of
this equation
\begin{eqnarray}
\left| v \frac{\partial U}{\partial y}\right| =|v|S\sim
\langle v^2\rangle^{1/2} S \le \langle u^2\rangle^{1/2} S\,.
\end{eqnarray}
In the last step above, we have used the fact that the
root-mean-square velocity in the streamwise direction is larger
than that in the shear direction, as has been found in all
numerical and physical experiments. The term on the right hand
side can be estimated roughly as $\langle u^2\rangle/{\ell}$
where the scale $\ell$ is characteristic of turbulent velocity
gradients, and can therefore be assumed to be of the order of the
Taylor microscale, $\lambda$. We then require
\begin{eqnarray}
\langle u^2 \rangle^{1/2}\,S\gg \frac{\langle u^2
\rangle}{\lambda}\,.
\end{eqnarray}
With $\epsilon=c_{\epsilon} u^{\prime 3}/L$ and
$L/\lambda=Re_L^{1/2}/\sqrt{10}$, and the constant
$c_{\epsilon}\sim {\cal O}(1)$ for sufficiently large
$R_{\lambda}$, we get
\begin{eqnarray}
\frac{S^{\ast}}{R_{\lambda}}\gg \frac{\sqrt{3}}{\sqrt{200}}
\approx \frac{1}{8}\,.
\end{eqnarray}
In reality, $c_{\epsilon}$ depends weakly on $S^*$. For example,
Sreenivasan \cite{sreeni1} has examined the data and concluded
that $c_{\epsilon} \approx c_o \exp(-\alpha S^*)$, where $\alpha
\approx 0.03$, is a good empirical fit. Since this dependence is
quite weak, we have taken $c_{\epsilon}$ to be a constant for
simplicity. Presumably, if Eq.~(14) holds, the effects of shear
will always be felt no matter how high the Reynolds number.

\subsection{Local isotropy limit}
At the other extreme is the case in which local isotropy can be
expected, {\it a priori}, to prevail. A suitable criterion (see
Corrsin \cite{corrsin}) is that a sufficiently large separation
should exist between the shear time scale, $S^{-1}$, and the
Kolmogorov time scale, $\tau_{\eta}=(\nu/\epsilon)^{1/2}$. This
can be written as
\begin{eqnarray}
S\tau_{\eta}\ll 1\,,
\end{eqnarray}
or as
\begin{eqnarray}
S\tau_{\eta}=S^{\ast}\sqrt{c_{\epsilon}}\sqrt[4]{\frac{3}{2}}Re_L^{-1/2}\,.
\end{eqnarray}
With $c_{\epsilon}\sim {\cal O}(1)$ and $R_{\lambda}=(20
Re_L/3)^{1/2}$ we obtain
\begin{eqnarray}
\frac{S^{\ast}}{R_{\lambda}}\ll \frac{\sqrt[4]{3}}{\sqrt[4]{200}} \approx
0.35 \,.
\end{eqnarray}
The implication is that local isotropy prevails for all
$S^*/R_{\lambda}$ substantially smaller than 0.35. For all other
conditions, one should expect that the magnitude of $S^*$ will
play some role in determining how high an $R_{\lambda}$ is
required for local isotropy to prevail. This explicit dependence
on shear has been noted for passive scalars by Sreenivasan and
Tavoularis;\cite{sreeni2} see their figures 2 and 3.

\section{The $R_{\lambda}$-$S^{\ast}$ phase diagram}
We now plot in Fig.~6 all available data on the skewness of the
transverse derivative, $\partial u/\partial y$, on a phase plane
consisting of $S^*$ and $R_{\lambda}$. The conventional
normalization factors in the definition of $S^*$ are the total
turbulent energy and its dissipation rate. This can be done quite
readily for the numerical data, but experiments usually provide
information only on the streamwise component of the turbulent
energy and on the energy dissipation estimate from the local
isotropy formula, $\epsilon=15\nu\langle(\partial u/\partial
x)^2\rangle$. The error made in this estimate for the energy
dissipation depends\cite{sreeni1} on the magnitudes of shear and
$R_{\lambda}$, but it appears to be a reasonable approximation for
the present conditions. We have recalculated for all numerical
data the energy dissipation rate as in experiments. It is clear
from the figure that there is no simple correlation between the
two parameters $S^*$ and $R_{\lambda}$.

We have replotted the same data in Fig.~7.  Different symbols
correspond now to different magnitudes of the skewness, as
indicated in the legend, and not to different experiments.
Superimposed are islands in shades of grey to obtain a rough idea
of the surface plot of $\tilde{f}(R_{\lambda}, S^{\ast})$. We used
an interpolation routine with local thin plate splines that can
reconstruct a surface from scattered data.\cite{Franke} The data
are sparse in most of the regions of the plane so any
surface-fitting routine will introduce some peculiarities. For
reasons explained above, we can expect that the derivative
skewness will become small in the local isotropy limit (lower
right corner) and that the values will grow above 0.8 in the large
shear case (upper left corner). Although the latter fact is not
reflected by the surface fit because the data points are absent
there and almost all data points are in the intermediate region,
we think that the surface plot is not unreasonable.

This plot offers additional perspectives. For instance, Fig.~1 is
just a projection of the data on to the $R_{\lambda}$ axis and
masks the fundamental effect of the applied shear (among other
effects). We have marked in this figure the trends of the data
sets of Warhaft and co-workers (parabolic solid line labelled by
GW/SW) and of Ferchichi and Tavoularis (straight line labelled by
FT), respectively. They show that the two experiments follow
different paths: while the Ferchichi-Tavoularis data run directly
down the `mountain range', the Shen-Warhaft data seem to run in a
kind of spiral around the `mountain', presumably resulting in a
slower decay of the skewness when projected on to the
$R_{\lambda}$ axis. Finally, the two limiting regimes of Sec.~III
are also plotted. The local isotropy limit is not reached for any
of the data. The large shear case is reached for very large
$S^{\ast}$. A threshold $S^{\ast}_c$ grows when $R_{\lambda}$
increases; alternatively, local isotropy requires larger
$R_{\lambda}$ if the shear parameter is larger.

\section{Concluding remarks}
Considerable attention has been paid recently to the fact that in
shear flows the moments of transverse velocity derivatives do not
vanish with Reynolds number as fast as was expected. We have
introduced two considerations for interpreting these observations,
invoking small-scale intermittency and the magnitude of the shear
parameter. These two effects work in combination with the Reynolds
numbers in determining the magnitudes of odd moments of velocity
derivatives. The fact that the odd moments, when normalized by an
appropriate power of the variance (a procedure steeped in studies
of Gaussian processes), decay more slowly than expected should not
be considered {\it a priori} as incontrovertible evidence against
local isotropy. We believe that the broader perspective of this
paper may explain some seeming contradictions that exist in the
literature.

The conclusions we draw in this paper would be more definitive if
the data spanned much higher Reynolds number range. This can be
done with adequate resolution only in atmospheric flows at
present. The existing measurements are in conformity with the
discussion here, but it is difficult to be definitive because of
the usual problems that often exist in atmospheric measurements.
On the other hand, it seems quite feasible in numerical
simulations to fix the Reynolds number and vary the shear
parameter, even though the Reynolds number range may be limited.
Such a study will tell us more about the restoration (or
otherwise) of small-scale isotropy.

In the recent past, various efforts have been made to understand
the effects of anisotropy through an SO(3) decomposition of
structure functions and other tensorial objects (see Ref.~29 for
the basic theoretical idea and Refs.~30-32 for implementations of
the idea and further references). The method offers a transparent
way of determining the degree of anisotropy in turbulent
statistics. The relation between that effort and the present
global picture needs to be explored.

\acknowledgments

We thank A.~Pumir and Z.~Warhaft for providing raw numbers for
some of their published figures, and acknowledge fruitful
discussions with A.~Bershadskii, S. Chen, J.~Davoudi, B.~Eckhardt,
R.J. Hill, S.~Kurien, S.B.~Pope, I. Procaccia (especially) and
Z.~Warhaft. J.S.~was supported by the Feodor-Lynen Fellowship
Program of the Alexander-von-Humboldt Foundation and Yale
University. The work was supported by the NSF grant DMR-95-29609.
Numerical calculations were done on a Cray T90 at the John von
Neumann-Institut f\"ur Computing and on the IBM Blue Horizon at
the San Diego Supercomputer Center. We are grateful to both
institutions for support.

\end{multicols}
\pagebreak

\begin{figure}
\begin{center}
\epsfig{file=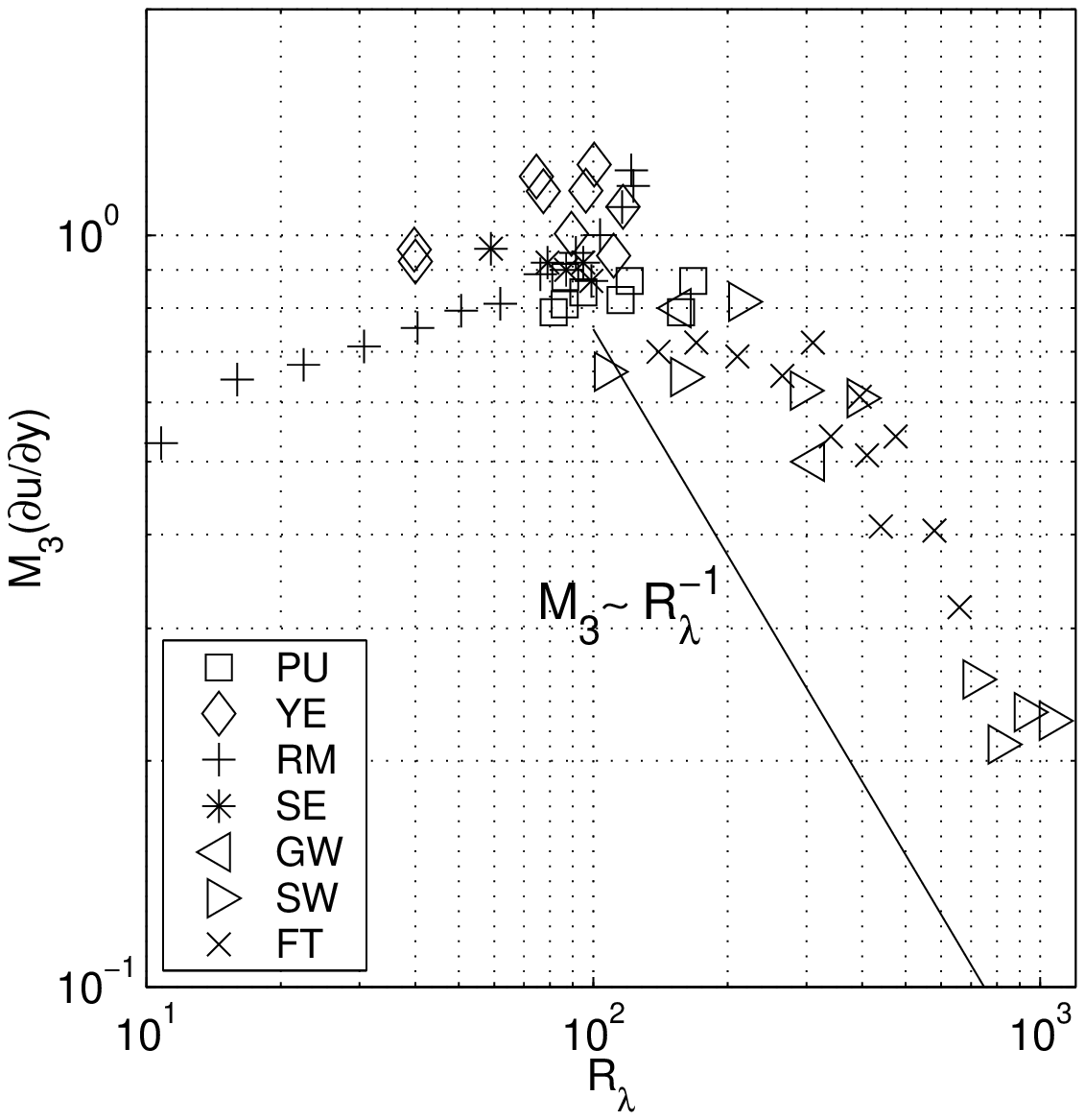,width=8cm}
\end{center}
\caption{The dependence of the skewness of the transverse velocity
derivative on the Taylor microscale Reynolds number. The solid
line indicates the $R_\lambda$ scaling expected to hold on the
basis of Lumley's dimensional considerations (see text). Notation
in the legend stands for: PU, Pumir [5]; YE, Yeung, for details of
whose shear flow simulations, see [12]; RM, Rogers {\it et al.}
[13]; SE, Schumacher and Eckhardt [6,7]; GW, Garg and Warhaft [2];
SW, Shen and Warhaft [4]; FT, Ferchichi and Tavoularis [3].}
\end{figure}
\begin{figure}
\begin{center}
\epsfig{file=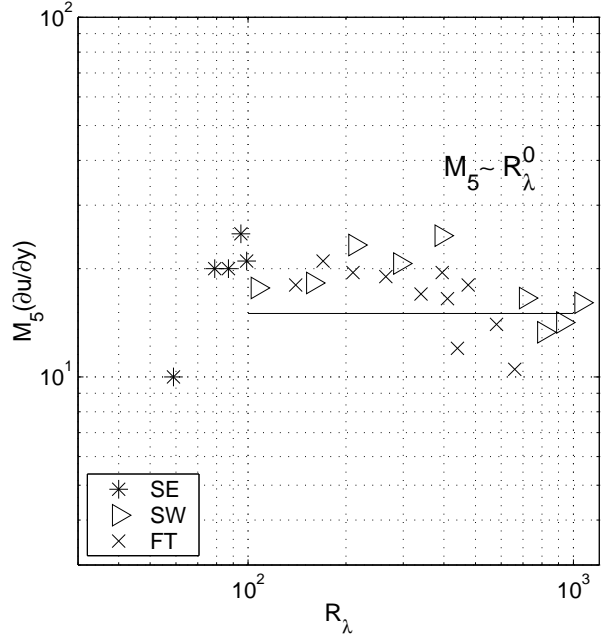,width=8cm}
\end{center}
\caption{The dependence of the hyperskewness of the transverse
velocity derivative on the Taylor microscale Reynolds number. The
notation in the legend is the same as in Fig.~1.}
\end{figure}
\begin{figure}
\begin{center}
\epsfig{file=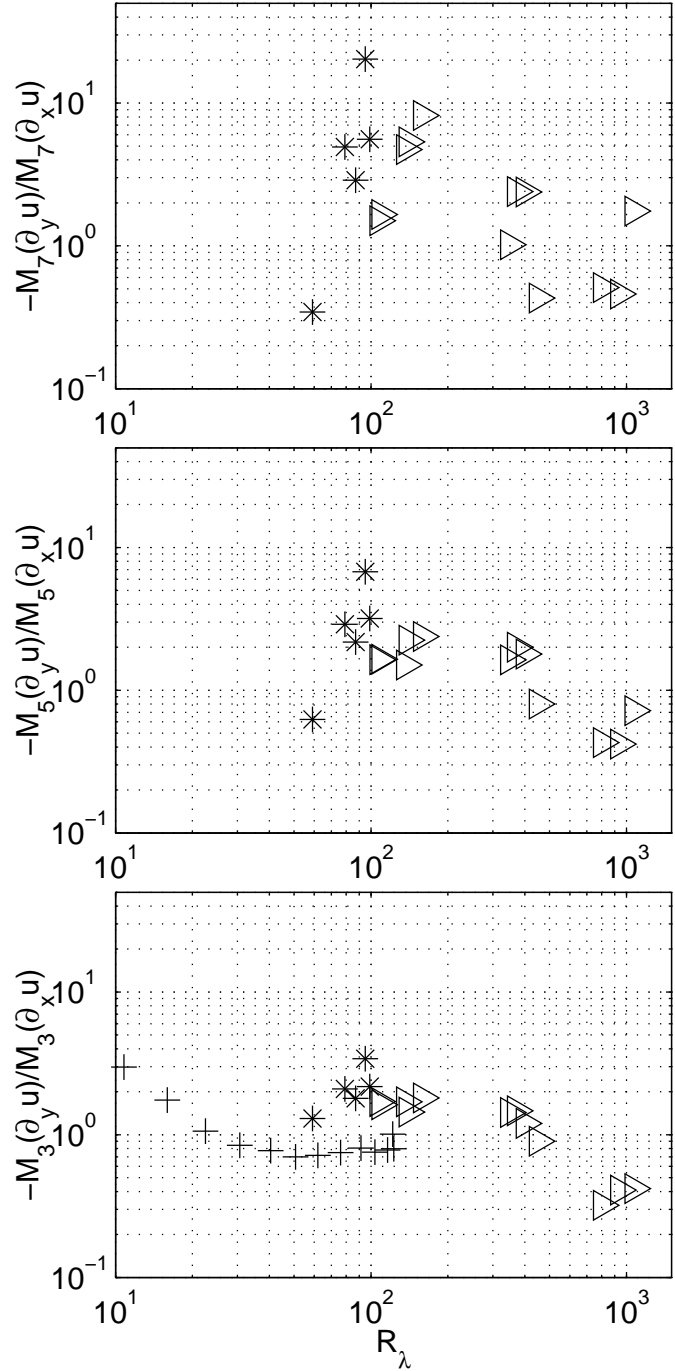,width=9cm}
\end{center}
\caption{The ratio of the normalized moments $M_{2n+1}(\partial
u/\partial y)$ and $M_{2n+1}(\partial u/\partial x)$ for $n$ = 1,
2 and 3 versus the Taylor microscale Reynolds number. Quantities
$M_{2n+1}$ are defined in Eq.~(1). The notation in the legend is
the same as in Fig.~1. Here and elsewhere, data from [4] have been
read from published graphs.}
\end{figure}
\begin{figure}
\begin{center}
\epsfig{file=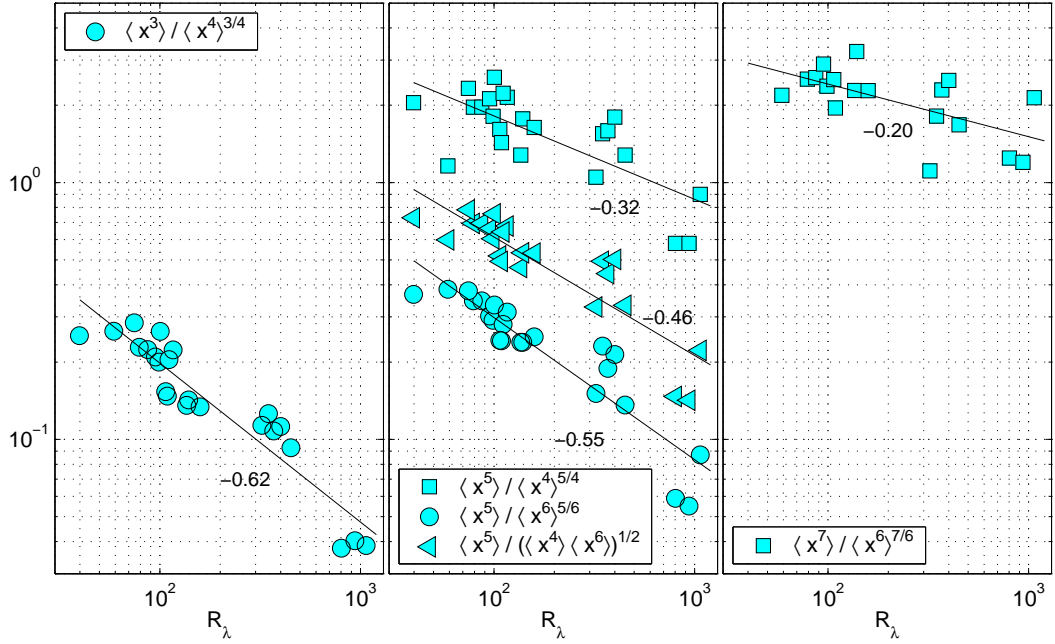,width=14cm}
\end{center}
\caption{Third, fifth, and  seventh order transverse derivative
moments with different normalizations, plotted against
$R_{\lambda}$. Squares are normalized by the moment of preceding
even order, circles by the succeeding even order. Triangles are
normalized by the geometric mean of both orders. Variable $x$ in
the legend stands for $\partial u/\partial y$. The data are from
Yeung [12], Schumacher/Eckhardt [6,7] and Shen/Warhaft [4]. Solid
lines with attached numbers indicate fits to possible algebraic
power laws. Each symbol corresponds to a different normalization,
and includes all the data just cited.}
\end{figure}
\begin{figure}
\begin{center}
\epsfig{file=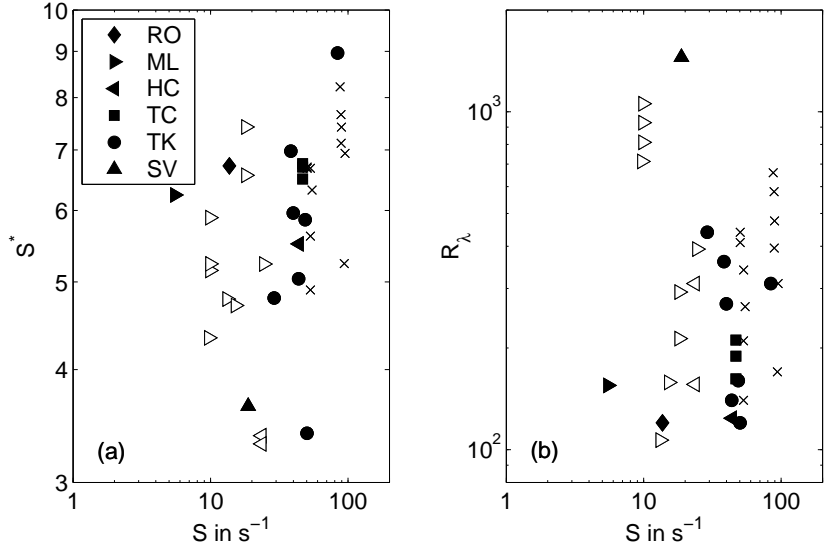,width=11cm}
\end{center}
\caption{ (a) The dependence of the shear parameter $S^{\ast}$ on
the shear rate $S$ for fixed kinematic viscosity, $\nu=\nu_{air}$.
(b) Dependence of the Taylor-microscale Reynolds number
$R_{\lambda}$ on $S$ for fixed $\nu$. Unfilled symbols are the
same as in Fig.~1. Additional shear flow experiments have been
included, though they did not focus on derivative moments
explicitly (filled symbols). The additions are: RO, Rose [17]; ML,
Mulhearn and Luxton [18]; HC, Harris, Graham, and Corrsin [19];
TC, Tavoularis and Corrsin [20]; TK, Tavoularis and Karnik [21];
SV, Saddoughi and Veeravalli [22].}
\end{figure}
\begin{figure}
\begin{center}
\epsfig{file=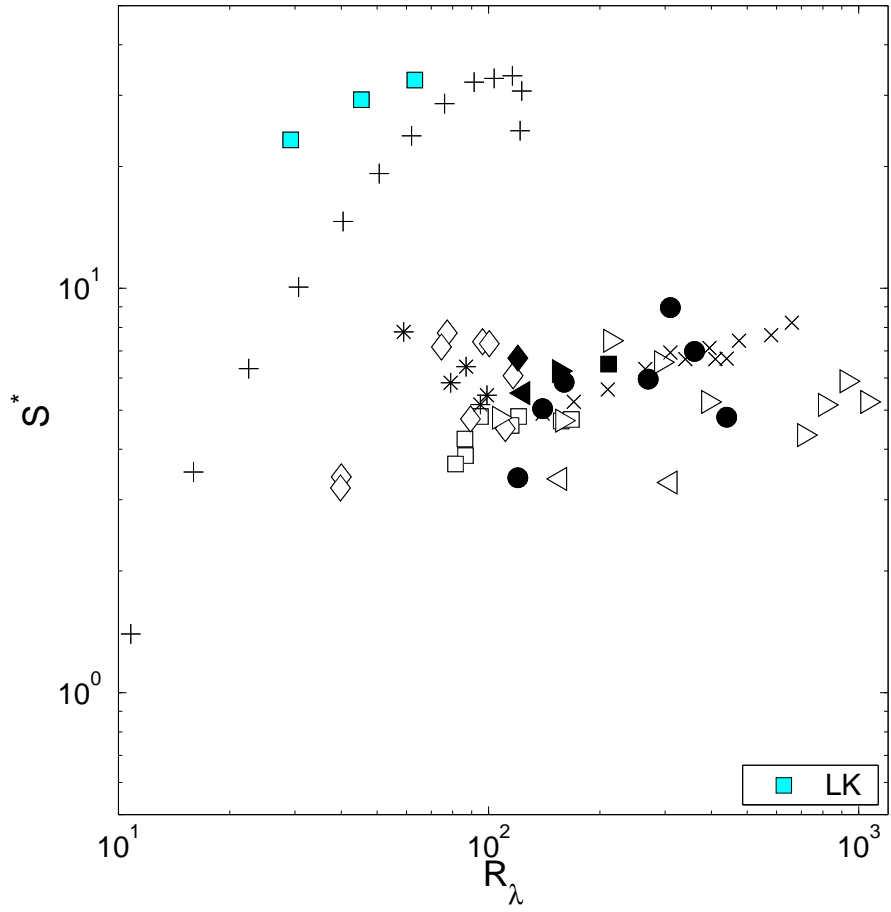,width=12cm}
\end{center}
\caption{Operating points of the homogeneous shear flows in the
$R_{\lambda}$-$S^{\ast}$ plane. All symbols are the same as in
Figs.~1 and 5. Data points LK stand for Lee {\it et al.} [26],
corresponding to additional numerical simulations at very high
shear rates.}
\end{figure}
\begin{figure}
\begin{center}
\epsfig{file=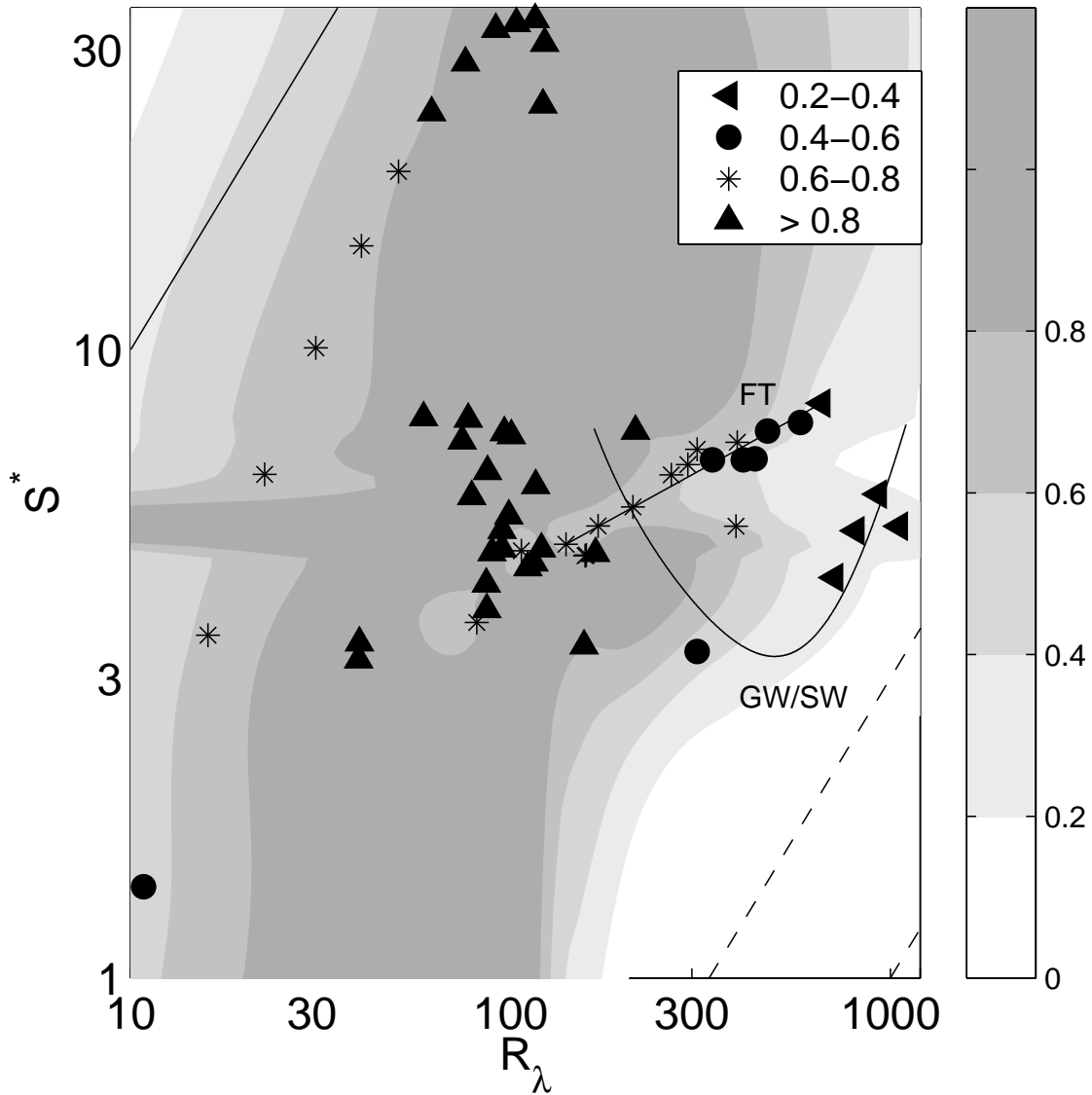,width=15cm}
\end{center}
\caption{Derivative skewness and its dependence of $R_{\lambda}$
and $S^{\ast}$. The solid straight line is for the large shear
limit $S^{\ast}/R_{\lambda} = 1$, the dashed lines are for local
isotropy limit $S^{\ast}/R_{\lambda}=0.003$ and 0.001 (left to
right). Underlying grey scales result from a surface fit. Trends
of the data from Ferchichi and Tavoularis (FT) and Warhaft and
co-workers (GW/SW) are indicated by solid lines. Different symbols
indicate different ranges of the skewness, and may represent data
from the same source. The data point below $S^{\ast}=3$ is taken
from the numerical experiments of Rogers {\it et al.} [13]}
\end{figure}
\end{document}